\documentclass[sigconf]{acmart}

\usepackage{tikz}
\usetikzlibrary{arrows.meta,positioning,calc,fit}
\usepackage{enumitem}
\usepackage{booktabs}

\newcommand{\nIEEE}{231}
\newcommand{\nACM}{512}
\newcommand{\nSCOPUS}{618}
\newcommand{\nWOS}{197}
\newcommand{\nMERGED}{1{,}246}
\newcommand{\nFT}{69}
\newcommand{\nINC}{44}
\newcommand{\nFINAL}{50}

\copyrightyear{2026}
\acmYear{2026}
\setcopyright{cc}
\setcctype{by}
\acmConference[ICMI Companion '26]{Companion of the INTERNATIONAL CONFERENCE ON MULTIMODAL INTERACTION}{October 05--09, 2026}{Napoli, Italy}
\acmBooktitle{Companion of the INTERNATIONAL CONFERENCE ON MULTIMODAL INTERACTION (ICMI Companion '26), October 05--09, 2026, Napoli, Italy}
\acmDOI{10.1145/3776591.3837048}
\acmISBN{979-8-4007-2319-3/2026/10}

\begin{document}

\title[Adaptive Human-AI Collaboration]{Adaptive Human-AI Collaboration: A Review of Multimodal Context Modeling, Uncertainty-Aware Intervention, and Longitudinal Co-Adaptation}

\author{Mahyar T. Moghaddam}
\email{mtmo@mmmi.sdu.dk}
\affiliation{%
  \institution{University of Southern Denmark}
  \city{Odense}
  \country{Denmark}}

\author{Mina Alipour}
\email{mial@mmmi.sdu.dk}
\affiliation{%
  \institution{University of Southern Denmark}
  \city{Odense}
  \country{Denmark}}
  
\renewcommand{\shortauthors}{Moghaddam et al.}

\begin{abstract}
Artificial intelligence is shifting from a static decision-support tool to an adaptive collaborator that must sense context, decide when and how to intervene, and improve through repeated interaction with humans individually and in groups. Yet meta-analytic evidence shows that human-AI combinations often fail to outperform the best of either partner alone, and the enabling literature remains fragmented across multimodal sensing, uncertainty quantification, reliance and delegation, facilitation, and teaming.
This paper reports a review of adaptive human-AI collaboration literature through a closed-loop lens. Following iterative identification, staged selection against explicit criteria, structured extraction, taxonomy-driven synthesis, and snowballing, we analyze 50 reviewed works. We contribute {\em i)} a taxonomy of multimodal context modeling, from individual states to collective states such as group engagement and participation equality; {\em ii)} a taxonomy of uncertainty-aware intervention, covering uncertainty sources, estimation and calibration mechanisms, an intervention repertoire that ranges from explanation modulation and deferral to group facilitation, and intervention policies; and {\em iii)} a taxonomy of longitudinal co-adaptation and synergy-oriented evaluation. We integrate the three taxonomies into MCAL, a dual-timescale Multimodal Co-Adaptation Loop reference model, and instantiate it on a mixed human-robot workspace, walking every stage of the loop through one concrete setting to show what each taxonomy cell holds in practice.
\end{abstract}


\maketitle
\emergencystretch=1.6em

\section{Introduction} \label{sec:intro}
AI systems increasingly act inside human activities rather than beside them: they draft, diagnose, drive, mediate meetings, and share factory floors. This shift turns a one-shot prediction problem into a \emph{collaboration} problem in which the AI must continuously answer three questions: \emph{what is going on} with the human(s), the task, and the environment; \emph{whether and how to act} given what it does not know; and \emph{how the partnership should change} over repeated interaction. The stakes of getting this loop right are documented at scale: a preregistered meta-analysis of 106 experiments found that, on average, human-AI combinations perform \emph{worse} than the best of human or AI alone, with losses concentrated in decision tasks [S48], and a reliability analysis of 52 clinical studies found that full complementarity is essentially never reached, while teaming mode and user expertise strongly moderate outcomes [S49]. The ``synergy gap'' is thus not a capability problem alone; it is a problem of how sensing, decision, and adaptation are closed into one loop.

Three research currents now make that loop technically plausible but remain siloed. First, multimodal sensing and foundation models can estimate rich human context, from pose and activity across six sensor types with a single modality-invariant encoder [S1] to grounded social reasoning over visual, verbal, and vocal cues [S3], and increasingly target \emph{collective} states such as group engagement across cultures and group sizes [S7, S8]. Second, uncertainty quantification has matured into deployable mechanisms with human-facing guarantees, from conformal prediction sets that measurably improve human decisions [S12] to calibrated ask-for-help triggering for LLM-driven robots [S13]. Third, a growing body of work treats adaptation as bidirectional and longitudinal: work on human-AI coevolution theorizes how people and AI can co-evolve through the data their interaction generates [S46], agents learn user preferences from naturally occurring edits across sessions [S32], and AI mediators iteratively move whole groups toward common ground [S39]. No existing review, however, synthesizes these currents as one closed loop (Section~\ref{sec:related}). The omission matters in multi-party settings: an adaptive AI peer or facilitator must estimate emergent, non-additive group states, act on them under uncertainty without dominating the interaction, and keep learning as the group learns it, a problem statement that spans all three currents at once and is served piecemeal by each.
This paper fills that gap with a literature review guided by three research questions:
\begin{itemize}[leftmargin=1.4em,itemsep=1pt,parsep=0pt]
\item \textbf{RQ1.} How do adaptive human-AI collaboration systems observe and model multimodal human, task, and social context, from individual to collective states?
\item \textbf{RQ2.} How is uncertainty estimated, calibrated, and used to select, shape, and time interventions?
\item \textbf{RQ3.} How do humans and AI adapt to each other over repeated interaction, and how is collaborative synergy evaluated?
\end{itemize}

We make four contributions. {\em 1)}~A taxonomy of \emph{multimodal context modeling} that spans modality portfolios, modeled constructs (activity, affect, competence, trust state, and collective states), fusion and temporal mechanisms, and interaction scope. {\em 2)}~A taxonomy of \emph{uncertainty-aware intervention} that separates uncertainty sources, estimation mechanisms, the intervention repertoire (including group-level facilitation and mediation), and intervention policies. {\em 3)}~A taxonomy of \emph{longitudinal co-adaptation} covering adaptation direction, timescale, mechanism, and synergy-oriented evaluation. {\em 4)}~We integrate the catalogs into \textbf{MCAL}, a dual-timescale \emph{Multimodal Co-Adaptation Loop} reference model rooted in feedback-loop architectures~\cite{kephart2003}, and instantiate it on a mixed human-robot workspace, walking every MCAL stage through one concrete setting to show that each stage decision is a taxonomy cell. 

\section{Related Reviews and Gap}\label{sec:related}
Several recent secondary studies border on our scope, and each leaves the closed loop uncovered. Vats et al.~\cite{vats2024} survey human-AI collaboration with large foundation models but organize it around human-guided development, design principles, governance, and applications rather than a runtime loop, and do not report a selection protocol. Fragiadakis et al.~\cite{fragiadakis2024} contribute an evaluation framework and metric decision tree; adaptivity appears as one evaluated factor, not the organizing mechanism. Wang et al.~\cite{wang2025hat} review adaptive human-agent teaming through team-process phases (formation, role development, team development, improvement), foregrounding teaming dynamics but not multimodal context modeling or uncertainty-driven intervention logic. Kumar et al.~\cite{kumar2025} map co-learning and co-adaptation terminology, agent types, and task domains in a scoping review, deliberately stopping short of a mechanism-level synthesis, and Iftikhar et al.~\cite{iftikhar2024} review human-agent team dynamics with a management orientation. On the systems side, reviews of machine learning \emph{inside} self-adaptive systems~\cite{gheibi2021} and of control-loop patterns \emph{around} AI-enabled systems~\cite{moghaddam2026controlloops} catalog feedback-loop architecture but do not treat humans as sensed, uncertain, adapting partners; reviews of robots in groups~\cite{sebo2020} and of human-in-the-loop cyber-physical systems~\cite{clemmensen2025} cover the social and socio-technical sides without the uncertainty-aware control core. Position work on socially intelligent agents likewise calls for, but does not synthesize, this integration~\cite{mathur2024advancing}.

The gap this review targets is therefore precise: \emph{no secondary study synthesizes adaptive human-AI collaboration as a closed loop connecting multimodal (including collective-state) context modeling, uncertainty-aware intervention, and longitudinal co-adaptation, with synergy as the evaluation target.}

\section{Method}\label{sec:method}
\providecommand{\nCAND}{124}
This paper reports a structured, concept-centric survey, conducted as an integrative review~\cite{snyder2019,torraco2005,pare2015} and organized around concepts rather than around authors~\cite{webster2002}. Its aim is coverage of the mechanism space rather than of the literature, so the method is built to make the catalog of mechanisms complete and traceable: identification, selection, extraction, and synthesis, with taxonomy development as the analytic core.
 
\textbf{Scope and criteria.} The three research questions of Section~\ref{sec:intro} fix the scope, and Table~\ref{tab:criteria} states the criteria applied before reading. A work qualifies when humans and AI interact toward a shared task with at least one adaptive mechanism in the loop, when it proposes, evaluates, or empirically characterizes such a mechanism, and when it addresses the interaction loop rather than offline model quality alone. Because the mechanisms of interest are also carried by benchmarks, challenges, and released models rather than by studies alone, the criteria admit any work that contributes to the loop, whatever its form.
 
\begin{table}[t]
\caption{Inclusion and exclusion criteria.}
\label{tab:criteria}
\footnotesize
\begin{tabular}{@{}p{0.27\columnwidth}p{0.65\columnwidth}@{}}
\toprule
\multicolumn{2}{@{}l}{\textbf{Inclusion criteria}}\\
\midrule
Subject (IC1) & Human(s) and AI interact toward a shared task, or the work contributes an enabling mechanism or resource directly relevant to such interaction, including context modeling, uncertainty estimation, intervention, delegation, facilitation, or longitudinal adaptation\\
Contribution (IC2) & Proposes, evaluates, benchmarks, or empirically characterizes mechanisms, systems, datasets, models, or evidence relevant to adaptive human-AI collaboration loops\\
Scope (IC3) & Maps to at least one stage of the collaboration loop. Purely offline model-quality work is excluded unless it is explicitly about collaboration-relevant constructs or is used as an enabling benchmark/resource\\
Type (IC4) & Peer-reviewed journal, conference, or workshop paper preferred. Preprints, challenge resources, and released benchmarks may be admitted when they are central enabling resources and no peer-reviewed equivalent is available\\
Language (IC5) & English\\
Access (IC6) & Full text available\\
\midrule
\multicolumn{2}{@{}l}{\textbf{Exclusion criteria}}\\
\midrule
EC1 & No human partner, no human-facing teaming context, and no collaboration-relevant mechanism\\
EC2 & Purely algorithmic or model-centric work with no explicit mapping to any collaboration-loop stage or construct\\
EC3 & Interaction is discussed, but no sensing, intervention, delegation, facilitation, or adaptation mechanism of interest is present\\
EC4 & Secondary studies are excluded from the primary evidence synthesis but may be included as contextual or motivating evidence where they provide aggregate empirical results (e.g., meta-analyses or reliability analyses).\\
EC5 & Non-English\\
EC6 & Full text not accessible\\
\bottomrule
\end{tabular}
\end{table}

\textbf{Identification.} Identification ran as an iterative search-and-read cycle rather than as a single retrieval event~\cite{webster2002,snyder2019}, through three parallel channels. \emph{Venue-driven browsing} covered the communities that carry this work: CHI, CSCW, ICMI, UMAP, HRI, and ACM Multimedia on the interaction side; NeurIPS, ICML, ICLR, CoRL, and the ACL venues on the modeling side; and the information-systems and general-science outlets that carry the evidence on complementarity. \emph{Concept-driven probing} over scholarly search engines operationalized the research questions into three groups, namely collaboration, adaptivity and loop mechanisms, and the multimodal, uncertainty, and co-adaptation pillars, and was re-run as new terminology surfaced rather than frozen into one string. \emph{Backward and forward snowballing}~\cite{wohlin2014,zhang2011} expanded from works judged central.
 
\textbf{Selection.} Selection made three passes over the pool of 124 candidate studies. Titles and abstracts were screened against Table~\ref{tab:criteria} and retained only when they met every inclusion and no exclusion criterion, giving \nFT{} candidates against 55 exclusions. Full-text assessment then gave \nINC{} works, with 25 exclusions recorded with reasons. Snowballing added six, four of them preprints or benchmarks admitted under IC4. The result is a corpus of \nFINAL{} reviewed works [S1 to S50]: 24 journal, 22 conference, and 4 workshop, preprint, or benchmark items; 33 appearing between 2023 and 2026; and, by interaction scope, 32 dyadic, 12 group-level, and 6 individual-adaptation works. Selection was split among the authors and cross-checked, with disagreements resolved by consensus.
 
\textbf{Stopping rule.} Since the design does not inherit exhaustiveness as its termination condition, it states one explicitly, and we adopt the ending conditions \cite{nickerson2013} that identification stopped when a full pass over new candidates produced no new dimension and no new leaf in any of the three taxonomies, when every dimension held at least one work, and when no dimension or leaf was merged, split, or renamed during that pass. The works added last extend the evidence behind existing cells without extending the catalog, which is the sense in which the survey is complete.
 
\textbf{Extraction.} Extraction was structured as a concept matrix~\cite{webster2002}, coding every work on the dimensions of the three taxonomies (Sections~\ref{sec:rq1} to \ref{sec:rq3}) plus venue, publication type, interaction scope, and evidence type, the last distinguishing controlled experiment, field study, benchmark, system-plus-study, and theory, so that the strength of evidence behind every cell stays traceable rather than implied. Table~\ref{tab:glance} previews the distributions that drive the synthesis: adaptation is coded for barely half the corpus and is mutual in only seven works; a mere eleven observe more than one episode; and although twelve operate at group level, only two close the loop from a rich collective-state model to a group-level intervention.
 
\begin{table}[t]
\caption{The corpus at a glance ($N=\nFINAL$).} 
\label{tab:glance}
\footnotesize
\begin{tabular}{@{}p{0.56\columnwidth}r@{}}
\toprule
Interaction scope: individual / dyad / group & 6 / 32 / 12\\
\midrule
Adaptation direction (CA1): AI$\rightarrow$human & 8\\
\quad human$\rightarrow$AI documented & 12\\
\quad mutual co-adaptation & 7\\
\quad none (single-shot design) & 23\\
Observation beyond a single episode (CA2) & 11\\
\midrule
Human-side uncertainty modeled at all (UI1) & 10\\
Intervention policy: learned / rule-threshold (UI4) & 9 / 6\\
\midrule
Group studies with a rich collective-state model & 5\\
Group studies with any group-level intervention & 8\\
Group studies with \emph{both} (closed collective loop) & \textbf{2}\\
\bottomrule
\end{tabular}
\end{table}
 
\textbf{Synthesis and taxonomy development.} Synthesis followed thematic synthesis~\cite{cruzes2011} in two stages, and taxonomy development the method of Nickerson et al.~\cite{nickerson2013}. Conceptual-to-empirical passes mapped works deductively onto the pillars implied by the research questions; empirical-to-conceptual passes refined the catalog inductively from the coded data, surfacing the AI-specific mechanisms the initial pillars did not anticipate, among them verbalized uncertainty, conformal help-triggering, facilitation policies, and learning from naturally occurring edits. The passes alternated until the ending conditions were met. A pilot round calibrated terminology across communities before coding began, reconciling reliance against trust calibration and personalization against co-adaptation, since one construct carries different names in different venues and a taxonomy must not inherit that fragmentation.

\section{RQ1: Multimodal Context Modeling}\label{sec:rq1}
\begin{figure*}
    \centering
    \includegraphics[width=\linewidth]{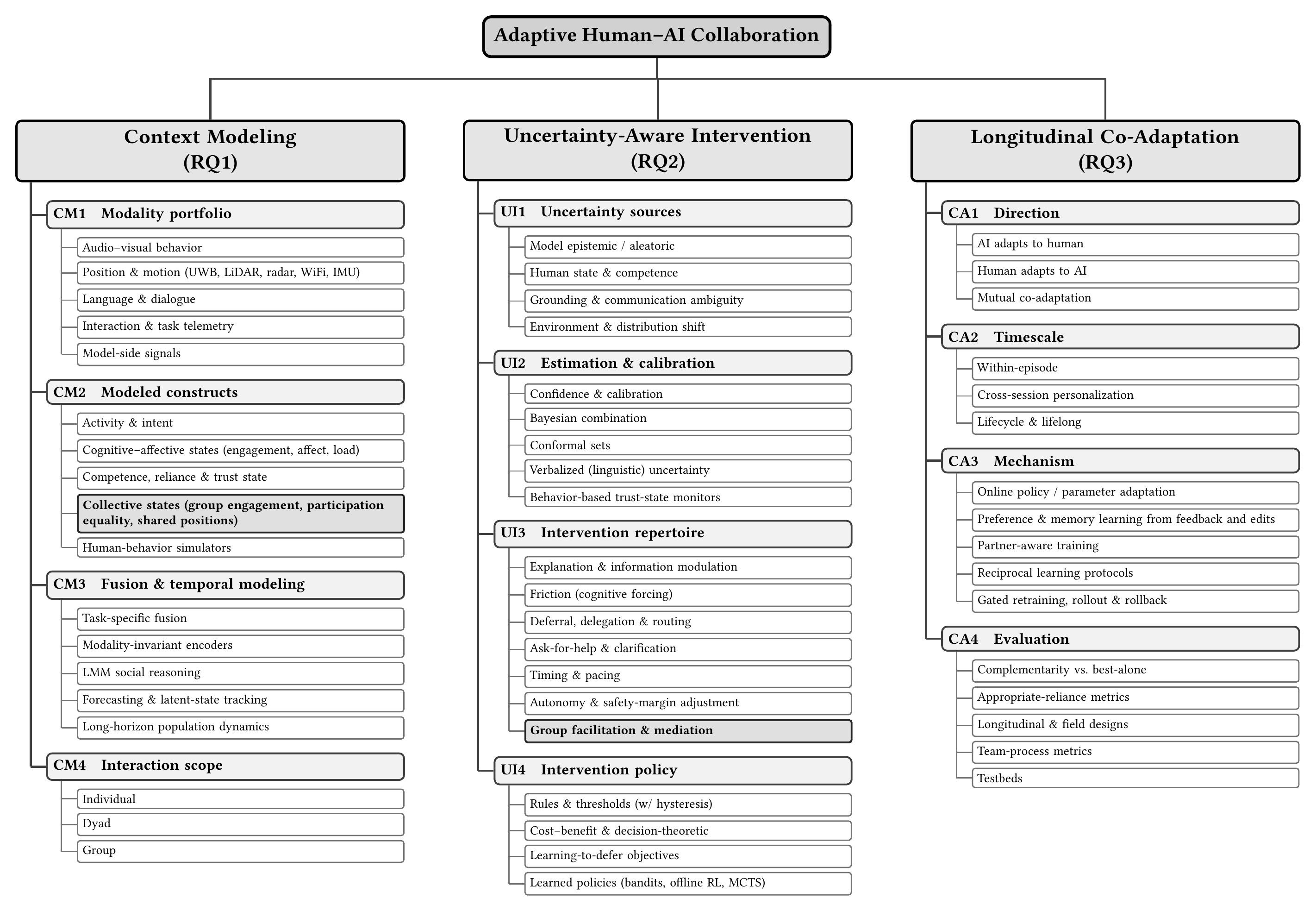}
\caption{Taxonomy tree for adaptive human-AI collaboration: the head branches into the three synthesized taxonomies: multimodal context modeling (RQ1), uncertainty-aware intervention (RQ2), and longitudinal co-adaptation (RQ3); each divided into dimensions and then into their sub-components. Shaded sub-components mark the collective-state entries most relevant to hybrid human-AI groups.}
\label{fig:taxonomy}
\end{figure*}

This section synthesizes how the reviewed works observe and model context, organized along three dimensions plus the cross-cutting interaction scope (Figure~\ref{fig:taxonomy}, left).

\textbf{Modality portfolio (CM1).}
The corpus spans five recurring input families: {\em i)}~\emph{audio-visual behavior}: speech, prosody, gaze, facial expression, gesture, and body pose, dominant in social-interaction sensing [S3, S4, S7, S8] and affect-driven adaptation [S9, S35]; {\em ii)}~\emph{position and motion sensing} beyond cameras: LiDAR, mmWave radar, WiFi CSI, and IMUs for pose and activity [S1], and speaking-activity or turn-level audio in groups [S40, S41]; {\em iii)}~\emph{language and dialogue}, the primary channel of LLM-era collaboration [S6, S13, S14, S31, S44]; {\em iv)}~\emph{interaction and task telemetry}: clicks, edits, reliance behavior, gameplay events, and workflow stages, which are cheap, privacy-lighter, and directly tied to the collaboration [S16, S30, S32, S36, S38]; and {\em v)}~\emph{model-side signals} such as confidences and behavior maps that become context \emph{about the AI} for the human [S11, S22]. Vision-free portfolios matter in privacy-constrained deployments; the modality-invariant encoder of [S1] shows that the portfolio can be decoupled from retraining, so sensors can be added or dropped at runtime, an enabler for adaptive loops that Sections~\ref{sec:mcal}-\ref{sec:case} exploit. Richer portfolios buy robustness and construct coverage at the cost of deployment, synchronization, and privacy burden; group settings sharpen the trade-off further, since always-on audio-video capture of bystanders raises consent questions that telemetry- or position-based portfolios largely avoid, and cross-cultural evaluation shows that portfolio choices interact with fairness of the resulting estimates [S8].

\textbf{Modeled constructs: from individual to collective states (CM2).}
What the signals are turned \emph{into} splits into four construct families. \emph{Activity and intent}: pose, action, trajectories, and task progress [S1, S10]. \emph{Cognitive-affective states}: engagement, affect, and workload, modeled continuously in [S7, S8, S9, S35]. \emph{Competence, reliance, and trust state}: the human's self-efficacy and metaknowledge [S18, S27], their moment-to-moment reliance behavior as an observable trust proxy [S16], the AI's own competence map communicated back to the human [S22], and predictive models of human behavior itself, up to foundation-model simulators of choice behavior [S5] and theory-of-mind-style inference tasks [S6], which also expose an action gap in current LLMs. \emph{Collective states}: group engagement across domains and cultures [S7, S8], participation equality and conversational dynamics [S36, S40, S41], discussion stage and stance structure [S38], and group (dis)agreement over positions [S39]. Crucially, the group-level works emphasize that a group state is not the sum of member states: [S7] annotates engagement at the interaction level across group sizes, [S40] shows a single agent's behavior propagating through human-human dynamics across thirty rounds of a team game, and [S41] demonstrates that even a peripheral, non-anthropomorphic object can be the sensing-and-actuation surface for a collective construct (participation balance). The construct inventory also distinguishes fast \emph{states} (momentary engagement, turn dynamics) from slow, trait-like properties (cohesion, shared mental models); the corpus measures the former far more often than the latter, which matters because interventions plausibly act on states while synergy accrues through traits. Yet only 12 of \nFINAL{} reviewed works model any collective construct, and fewer still feed it back into control (Section~\ref{sec:rq2}).

\textbf{Fusion and temporal modeling (CM3).}
Mechanisms range from task-specific multimodal fusion with cross-domain generalization pressure [S7, S8], through modality-invariant transformer fusion [S1], to foundation-model reasoning that fuses modalities implicitly and must ground its inferences in evidence [S2, S3]. Temporality appears as short-horizon forecasting of human motion and intent for proactive assistance [S10], latent-state tracking over dialogue [S44], and long-horizon population dynamics [S46]. Two evaluation results discipline this space: holistic benchmarking shows multimodal skills are uneven across interaction types and use cases [S2], and grounded social reasoning remains far from human traces even for frontier models [S3], so context models must expose their uncertainty rather than presume competence, which is the bridge to RQ2. Temporal granularity is a design decision: frame-level estimates feed reactive control, short-horizon forecasts enable proactive assistance and safety margins [S10], and session- or population-level dynamics inform the slow adaptation loop [S46], mirroring, on the sensing side, the dual timescales that Section~\ref{sec:mcal} makes explicit.

\emph{Synthesis.} Sensing is the most mature pillar, but it is largely built and evaluated \emph{open-loop}: benchmarks reward estimation accuracy, not downstream intervention value, and collective-state estimation in particular is rarely wired to any actuator.

\section{RQ2: Uncertainty-Aware Intervention}\label{sec:rq2}
Figure~\ref{fig:taxonomy} (center) organizes the second pillar into sources, estimation, repertoire, and policy.

\textbf{Uncertainty sources (UI1).}
Four sources recur. \emph{Model uncertainty} over predictions and plans [S11, S12, S17]. \emph{Human-state and competence uncertainty}: whether the person is over- or under-relying [S16], whether they possess the metaknowledge to delegate well, which they systematically do not [S27], and how traits such as self-efficacy shape delegation [S18]. \emph{Grounding and communication ambiguity}: which instruction interpretation is intended [S13, S44], and how hedges in natural language are produced and read [S14, S15]. \emph{Environment and distribution shift}, from cross-domain engagement estimation [S7, S8] to preference drift induced by the loop itself [S46]. A structural finding: the machinery for the first source is far ahead of the other three. Only 10 of \nFINAL{} reviewed works model human-side uncertainty at all, and mostly through coarse traits, self-reports, or reliance behavior rather than online multimodal estimation, even though the sensing stack of RQ1 produces exactly the signals such estimators would need.

\textbf{Estimation and calibration (UI2).}
Deployable mechanisms include calibrated confidence and its Bayesian combination with human confidence, which yields provable conditions for complementarity even between unequal partners [S11]; conformal prediction, which converts scores into coverage-guaranteed sets [S12] and into statistically calibrated triggers for asking humans for help [S13]; behavior-based monitors that detect miscalibrated trust from reliance patterns [S16]; and \emph{verbalized} uncertainty for LLMs: deployed models rarely hedge unprompted, and when prompted toward expressions of confidence they skew strongly overconfident, with human-subject experiments tying these miscalibrated verbal signals directly to inflated reliance and downstream error [S14]; input-side epistemic markers likewise shift model behavior itself, so linguistic uncertainty is simultaneously an estimation channel and an intervention surface [S15]. Cascade analysis shows plain confidence thresholds suffice only under restrictive conditions, motivating learned deferral criteria [S17].

\textbf{Intervention repertoire (UI3).}
The corpus yields seven recurring intervention families. (1)~\emph{Explanation and information modulation}: explanations increase acceptance more than complementarity [S19] and reduce over-reliance only when they make verification cheap relative to the task [S21]; global behavior descriptions improve selective reliance [S22]; the very representation of AI output (labels, probabilities, retrieval-based evidence) interacts with user expertise, with less experienced users gaining most and representation choice modulating who benefits [S34]. Information modulation is therefore not one lever but a family: what to show, how to frame it, and whether to show anything at all. (2)~\emph{Friction}: cognitive-forcing designs cut over-reliance at a subjective-preference cost [S20], with process evidence on how professionals actually integrate advice [S23]. (3)~\emph{Deferral, delegation, and routing}: from learning-to-defer [S24-S26] to cascades [S17] and bidirectional human-AI delegation, where the AI delegates well but humans do not [S27]. (4)~\emph{Asking for help and clarification}: calibrated help requests [S13], proactive clarification from joint belief states [S44], and learned hand-back of control in LLM agents [S31]. (5)~\emph{Timing and pacing}: when advice arrives changes anchoring and outcomes [S50], and across 52 clinical studies the simultaneous teaming mode outperforms the legally favored sequential one, with senior clinicians gaining almost nothing from sequential advice [S49], evidence that timing is a policy variable with regulatory implications, not a UI detail. (6)~\emph{Autonomy and safety-margin adjustment}: proactive robot assistance and task re-allocation as human intent becomes predictable [S10]. (7)~\emph{Group facilitation and mediation}: time-keeping, prompting quiet members, and summarizing in chat groups [S36]; structured deliberation moderation [S37]; stage-aware support for a human moderator [S38]; implicit participation balancing through a peripheral robotic object [S41]; social signals that reshape human-human dynamics [S40]; and iterative consensus-statement mediation that outperforms human mediators at scale [S39]. Family (7) is the direct intervention-side counterpart of collective-state sensing, yet the two sides barely meet: of the twelve group-level studies, five build a rich collective-state model and eight intervene at group level, but only two (stage-aware moderator support [S38] and iterative consensus mediation [S39]) do both. Facilitation otherwise acts on \emph{rules over shallow signals} (talk time, message counts) rather than on the collective-state estimators of RQ1, leaving the closed collective loop almost empty.

\textbf{Intervention policy (UI4).}
Policies range from thresholds and rules, sufficient for scripted facilitation [S36, S41] and trust-calibration cueing [S16], through cost-benefit accounts of when humans will engage with an intervention [S21], to formally derived learning-to-defer objectives [S24-S26] and learned policies: model-based planning over predictive user models for conservative UI adaptation [S30], bandit-style preference inference [S32], RL-tuned involvement policies for LLM agents [S31], offline RL that selects \emph{which} assistance type to give \emph{whom, when}, optimizing accuracy or human learning [S29], and hybrid designs that sandbox an LLM's action suggestions inside an auditable RL controller [S35]. Across the corpus, nine works learn their intervention policy while six rely on rules or thresholds; the rest fix a single intervention by design. Rules are inspectable but brittle under shift; learned policies personalize but raise assurance questions (what a policy may do, to whom, under which uncertainty) that almost no reviewed work addresses head-on, and that the lifecycle-governance literature answers only for models, not for interventions~\cite{moghaddam2026controlloops}.

\emph{Synthesis.} The field has calibrated \emph{whether to act} (defer, ask, abstain) better than \emph{how to act}: intervention selection, shaping, and timing are mostly studied one lever at a time, and only the RL-based strands [S29-S31, S35] treat the repertoire as a decision space.

\section{RQ3: Longitudinal Co-Adaptation}\label{sec:rq3}
\textbf{Direction (CA1).}
Adaptation direction is coded for only 27 of \nFINAL{} reviewed works; the remaining 23 are single-shot designs in which nobody adapts. Where present, it is mostly one-way. AI$\rightarrow$human (8 reviewed works): personalization from edits [S32], partner-specific deferral [S25, S26], planner adaptation to user costs [S30], and affect-driven game adaptation [S35]. Human$\rightarrow$AI adaptation (12 reviewed works) is documented as mental-model formation from behavior descriptions [S22], expertise-dependent gains [S34, S45, S49], and a homogenization risk: repeated exposure to AI advice can erode the unique knowledge that made the human complementary [S28]. Genuinely \emph{mutual} adaptation remains rare and recent (7 reviewed works): interaction-pattern catalogs of human-robot co-learning [S42], reciprocal human-machine learning protocols in operational classification [S43], bidirectional belief co-adaptation with proactive clarification [S44], group positions and mediator statements co-evolving over rounds [S39], and the population-scale coevolution frame [S46].

\textbf{Timescale (CA2).}
Adaptation happens \emph{within an episode} (most of UI3/UI4), \emph{across sessions} via preference memories and mental models [S22, S32, S43], and over the \emph{lifecycle}: sustained field use with heterogeneous gains [S45], long-run knowledge effects [S28], and societal-scale loops [S46]. Only eleven reviewed works observe interaction beyond a single episode; the mismatch with CA1 is direct: mutuality, deskilling, and homogenization are properties of trajectories, and cannot be measured in the single-shot designs that dominate the corpus.

\textbf{Mechanism (CA3).}
Mechanisms include online policy or parameter adaptation [S29-S31, S35]; preference and memory learning from naturally occurring feedback such as edits [S32]; training regimes that bake adaptability in by exposing agents to adapting partners or partner samples [S25, S26]; explicit reciprocal-learning protocols where both sides teach [S43]; and lifecycle machinery (drift detection, validation-gated retraining, staged rollout, rollback) which the self-adaptive-systems literature treats as first-class~\cite{gheibi2021,moghaddam2026controlloops} but which remains largely implicit in this corpus: adaptation mechanisms are proposed, yet how their updates are validated, activated, and revoked in deployment is rarely specified (see Sections~\ref{sec:mcal}-\ref{sec:case}).

\textbf{Evaluation and the synergy target (CA4).}
The corpus converges on complementarity as the demanding but right bar [S11, S47], and the aggregate answer is sobering: on average no [S48], and in 52 clinical studies never fully, with teaming mode and expertise as the strongest levers [S49]. Explanations alone do not close the gap [S19]; adaptive, person- and moment-targeted support can, in places [S29]. Appropriate-reliance metrics [S20-S22], process measures and testbeds for collaborative behavior [S33], field designs [S45], and reasoning-trace or cross-domain benchmarks on the sensing side [S2, S3, S7, S8] round out the measurement toolbox. What is missing is alignment: sensing benchmarks score estimation, intervention studies score single decisions, and almost nothing scores the \emph{loop}, whether closed-loop adaptation improves synergy over time. A loop-level benchmark would need at minimum: repeated episodes with the same partners, per-episode complementarity referenced to the best-alone baseline, appropriate-reliance trajectories rather than endpoints, and (in group settings) collective-state trajectories with individual outcomes; testbeds like [S33] and the cross-domain group corpora of [S7, S8] are usable starting substrates.

\section{MCAL: A Dual-Timescale Reference Model}\label{sec:mcal}
\begin{figure}
    \centering
    \includegraphics[width=\linewidth]{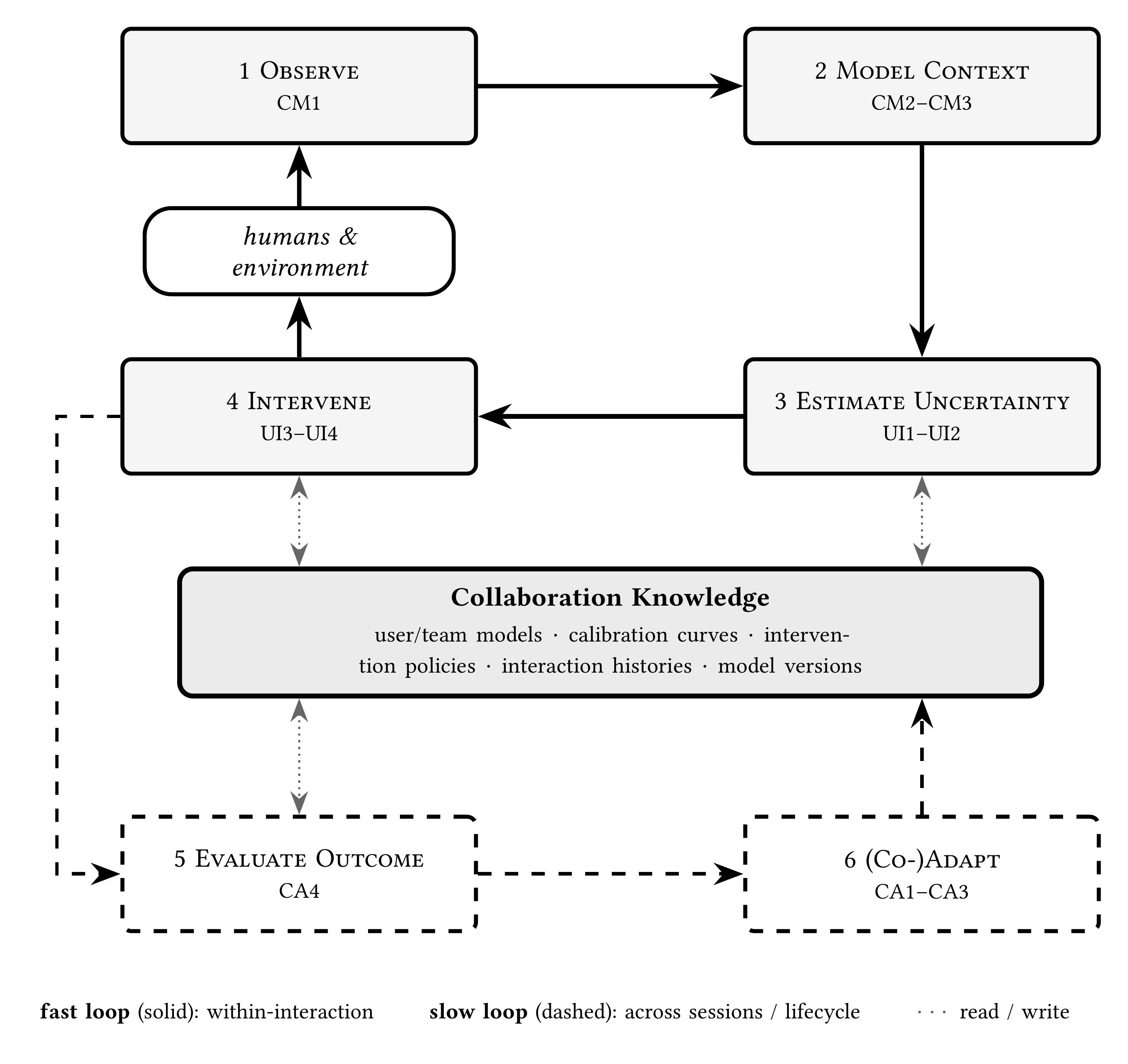}
\caption{MCAL, the Multimodal Co-Adaptation Loop: a fast interaction loop (solid) and a slow adaptation loop (dashed) over shared Collaboration Knowledge. Stage annotations reference the taxonomy dimensions of Figure~\ref{fig:taxonomy}; the ``co'' in stage 6 covers AI-side updates \emph{and} the human's learning.}
\label{fig:mcal}
\end{figure}

The three catalogs compose naturally into a closed loop. We synthesize them as \textbf{MCAL} (\emph{Multimodal Co-Adaptation Loop}, Figure~\ref{fig:mcal}): a fast \emph{interaction loop}: \textsc{Observe} (CM1) $\rightarrow$ \textsc{Model Context} (CM2-CM3) $\rightarrow$ \textsc{Estimate Uncertainty} (UI1-UI2) $\rightarrow$ \textsc{Intervene} (UI3-UI4); and a slow \emph{adaptation loop}: \textsc{Evaluate Outcome} (CA4) $\rightarrow$ \textsc{(Co-)Adapt} (CA1-CA3), both reading and writing a shared \emph{Collaboration Knowledge} substrate: user and team models, calibration curves, intervention policies, interaction histories, and model versions. The design descends from feedback-loop architectures for self-managing systems~\cite{kephart2003} and mixed-initiative interaction~\cite{horvitz1999}: \textsc{Observe}/\textsc{Model} refine \emph{Monitor-Analyze}, \textsc{Estimate}/\textsc{Intervene} refine \emph{Plan-Execute} with the human explicitly inside the plant, and the adaptation loop makes the lifecycle (retraining, recalibration, policy updates, and the \emph{human's} own learning) a first-class, governed concern rather than a side effect. To make the model concrete, consider an AI facilitator in a hybrid brainstorming meeting. \textsc{Observe} ingests audio, video, and chat telemetry (CM1); \textsc{Model Context} maintains per-member engagement and a group-level participation-equality and agreement estimate [S7, S36, S39]; \textsc{Estimate Uncertainty} attaches calibrated confidence to both (is the apparent disengagement of one member a sensing artifact, a cultural-display difference [S8], or real?) and tracks its own grounding of the discussion [S3]; \textsc{Intervene} then selects from the repertoire: stay silent, prompt a quiet member, restructure the agenda, propose a candidate consensus statement, or ask the group a clarifying question, with conformal-style triggering so that intrusive actions fire only under sufficient evidence [S12, S13]; \textsc{Evaluate Outcome} logs participation shifts, agreement movement, and member feedback; and \textsc{(Co-)Adapt} updates the facilitator's policy and per-team models across meetings while monitoring how the \emph{team} adapts to being facilitated [S28, S43]. Every stage decision above is a taxonomy cell, which is what makes the catalogs actionable rather than descriptive.

Three MCAL properties follow directly from the evidence: {\em i)}~\emph{uncertainty is the interface} between sensing and acting, interventions should be conditioned on calibrated uncertainty about model, human, and grounding, not on point estimates [S11-S17]; {\em ii)}~\emph{the repertoire is a decision space}, policies should select among explanation, friction, deferral, asking, timing, autonomy shifts, and facilitation [S29, S35], including at group level [S36-S41]; and {\em iii)}~\emph{both partners are plants and controllers}, evaluation must therefore be longitudinal and synergy-referenced [S43, S46-S49].

\section{MCAL in a Mixed Human-Robot Workspace}\label{sec:case}
Mixed human-robot workspaces, in which mobile robots and workers share a floor with no physical separation and human motion is the dominant uncertainty in the robot's planning problem, are a natural proving ground for a closed collaboration loop~\cite{moghaddam2026pitl}. We walk MCAL through one aisle of such a floor. The aisle carries three human roles: line operators who work at fixed stations and leave them only to fetch totes, one of them in their second week; a maintenance technician who moves between zones on no fixed route, because where they go next depends on a fault they have not yet described; and a shift supervisor who watches four aisles from a tablet and can be called. Two robots ferry parts from the store, and one performs a handover at a station every few minutes. Table~\ref{tab:case} collects the mapping.
 
\begin{table}[t]
\caption{MCAL walked through the mixed human-robot aisle of Section~\ref{sec:case}, with the taxonomy cells each stage draws on.}
\label{tab:case}
\footnotesize
\begin{tabular}{@{}p{0.19\columnwidth}p{0.735\columnwidth}@{}}
\toprule
\textbf{MCAL stage} & \textbf{What the aisle instantiates} \\
\midrule
1 Observe & UWB badges, robot poses and planned paths, station telemetry, opt-in wearables, a directional microphone at the handover; no vision, since visitors cross the aisle (CM1) \\
2 Model context & Working, crossing, or chasing a fault; load and skill from cycle times and andon pulls; reliance from yielding; congestion, contention, and wait-time balance as states no individual holds (CM2, CM3) \\
3 Estimate uncertainty & Conformal occupancy per horizon; a thin calibration curve for a two-week-old operator model; ambiguity over who ``wait'' addresses; shift and layout drift (UI1 all four sources; UI2) \\
4 Intervene & Inflate the envelope, project the intended path, ask one question, hand the segment back, explain to the novice and stay silent with the veteran, sequence two robots out of one segment (UI3, six of seven families); separation by rule with hysteresis, the rest learned (UI4) \\
5 Evaluate outcome & Complementarity against no-robot and full-autonomy baselines; reliance trajectories; congestion and waiting across people and robots; the novice's cycle time (CA4) \\
6 (Co-)adapt & Envelope widened around one technician within the shift; pacing learned from overrides across shifts; complacency monitored across months; shadow test, gate, and rollback before activation (CA1 mutual; CA2 all timescales; CA3) \\
\bottomrule
\end{tabular}
\end{table}
 
\textbf{1 Observe (CM1).} Badge-mounted ultra-wideband tags give positions for people and robots; the robots add their own poses and planned paths, which are model-side signals about the AI rather than about the human; station terminals emit task telemetry, namely cycle starts, part scans, and andon pulls; opt-in wearables give a coarse effort proxy; and a directional microphone captures speech addressed to a robot. Vision is absent by choice, since contractors and visitors cross this aisle and have consented to nothing.
 
\textbf{2 Model context (CM2, CM3).} Positions become activity and intent: an operator at a station is working, the same operator carrying a tote is crossing, and the technician is neither. Task telemetry becomes competence and load, since cycle times, rework, and andon pulls estimate per-operator workload and skill, and the second-week operator is modeled as such rather than as an average worker. Reliance is observable in behavior: an operator who has stopped glancing up at an approaching robot is habituated or complacent, and the two differ only in what they do when the robot behaves unusually. The aisle also holds states no individual holds: three people and two robots converging on one segment is congestion, a queue at the handover is contention, and the ratio of robot waiting to human waiting is a property of the arrangement rather than of any agent in it. Fusion runs at two horizons, a tracker for the next seconds and a shift-level model of who is where and how loaded.
 
\textbf{3 Estimate uncertainty (UI1, UI2).} All four sources are live at once. The trajectory forecast carries model uncertainty, wide for the technician and narrow for a stationary operator. The human-state estimate carries its own, since the wearable reports high load for the second-week operator while that operator's calibration curve rests on two weeks of data and says so. Grounding is ambiguous when someone says ``wait'' near the handover, because the word may address the robot, a colleague, or nobody; and the arrangement drifts, because the night shift routes differently and a moved pallet changes every path. The estimators match the sources: conformal occupancy regions per horizon rather than point predictions, a monitor comparing an operator's yielding against their own baseline, and verbalized hedging when the robot speaks, so that ``I think you asked me to wait'' reports a belief and not a fact.
 
\textbf{4 Intervene (UI3, UI4).} Clearance from the predicted human position is a fixed margin plus tracking error plus forecast error at that horizon, so the envelope inflates exactly when the model knows less and the robot yields room rather than acting on a poor estimate at unchanged confidence. Above that floor the choices are social. When the occupancy region grows, the robot makes its intent legible by projecting its path on the floor. When grounding is ambiguous, it asks one clarifying question instead of guessing. When the technician's intent is unreadable, it hands the segment back and requests human control. It explains its next move to the second-week operator and says nothing to the veteran, for whom the same explanation is noise. At aisle level the intervention is facilitation: the two robots are sequenced so they never enter one segment together, and a robot near the queue holds position rather than adding to contention. Policy differs by cell; separation is a rule with hysteresis and dwell times, so the envelope cannot oscillate and an auditor can read it; explanation, asking, and sequencing are learned, because their value depends on who is present and what happened last time.
 
\textbf{5 Evaluate outcome (CA4).} The unit of evaluation is the aisle over a shift, not the forecast. Beside safety margin and throughput, the loop logs whether the team beat both a no-robot and a fully autonomous baseline, whether operators' yielding tracked the robot's actual reliability or drifted toward blind trust, how congestion and waiting distributed across people and robots, and whether the second-week operator got faster. Forecast accuracy appears nowhere in that list, as it is an input to synergy rather than a measure of it.
 
\textbf{6 (Co-)adapt (CA1, CA2, CA3).} Within the shift the policy adapts online, widening the envelope around the technician who cuts corners. Across shifts it learns from overrides, since an operator who repeatedly waves a robot aside at the handover is stating a preference about pacing rather than generating noise. Over months the humans adapt too, and not always well: operators learn where robots yield and begin to depend on it, so the loop monitors the complacency it is itself producing. Every update is governed, since a new forecaster or policy is shadow-tested against the live one, passes a validation gate before atomic activation, and stays revocable, which matters more here than for a model alone, because the policy can now choose to speak to a person or to claim an aisle.
 
\textbf{Collaboration Knowledge.} Both loops read and write one substrate: per-operator competence and reliance models, per-aisle congestion history, calibration curves per horizon and zone, the intervention policy and its version, and the episode log the slow loop learns from. Nothing in the fast loop is personalized without something the slow loop wrote, and nothing the slow loop learns is trusted without the fast loop's telemetry.
 
\textbf{Synthesis.} No component above waits on an unavailable technology, since vision-free multimodal sensing, group-state estimation, conformal triggering, facilitation repertoires, and gated lifecycle machinery are all present in the corpus (Sections~\ref{sec:rq1} to \ref{sec:rq3}). What the example asks for, and what Table~\ref{tab:glance} says the field rarely delivers, is that they be wired into one loop: an estimator for the human and not only for the model, a collective state that reaches an actuator, an evaluation referenced to the best partner alone, and adaptation that runs both ways and is revocable when it goes wrong. Every stage above is a taxonomy cell and every arrow between stages is an interface whose currency is uncertainty, so the distance between this aisle and a deployed one is the distance the review measures.

\section{Discussion and Research Agenda}\label{sec:disc}
Five cross-cutting findings emerge. \textbf{(1) Two half-loops, rarely closed together.} Sensing-rich systems (RQ1) seldom calibrate or select interventions; intervention-rich systems (RQ2) run on thin context. Closing Observe$\rightarrow$Intervene end-to-end, with uncertainty as the interface, is the field's most direct path to synergy. \textbf{(2) The missing estimator is the human.} Model uncertainty is well served; online estimates of human state, competence, and reliance are proxied at best [S16, S18, S27]. Multimodal human sensing [S1, S7-S9] and behavior simulators [S5] are ready inputs for such estimators. 
\textbf{(3) Collective states are sensed but not steered.} Group engagement, cohesion-adjacent constructs, and participation are increasingly measurable [S7, S8], while facilitation systems still act on shallow signals [S36, S41]; wiring collective-state estimators to facilitation policies (with uncertainty-aware triggering) is a concrete agenda item, for which AI mediation of group positions [S39] shows the ceiling is high. \textbf{(4) Co-adaptation is asymmetric and short-horizon.} Mutual, multi-session evidence exists [S42-S44] but is scarce; longitudinal designs and shared testbeds [S33] are prerequisites for measuring it, and the homogenization risk [S28] warns that human-side adaptation is not automatically benign. \textbf{(5) Assurance of adaptive intervention is open.} Learned policies over interventions [S29, S35] raise the same drift, gating, and rollback questions that lifecycle-governed systems answer for models~\cite{moghaddam2026controlloops}; MCAL's knowledge substrate is the natural place to make intervention policies auditable, versioned, and revocable. Each finding doubles as an agenda item. \textbf{A1: Human- and group-state estimators with calibrated uncertainty:} treat engagement, workload, competence, and reliance as first-class estimands with conformal or Bayesian error bars, reusing the multimodal stack of RQ1. \textbf{A2: Closed-loop collective-state facilitation:} connect group-state estimators to facilitation repertoires under uncertainty-aware triggering, evaluated on participation, agreement, and cohesion trajectories, the near-empty cell of Table~\ref{tab:glance}. \textbf{A3: Loop-level, synergy-referenced longitudinal benchmarks:} multi-session protocols where the unit of evaluation is the adapting team, not a decision. \textbf{A4: Assurance for adaptive interventions:} versioned, auditable, revocable intervention policies with explicit authority envelopes, extending model-lifecycle governance to the intervention layer. \textbf{A5: Co-adaptation by design:} interaction protocols in which the human's learning is an optimization target and the homogenization risk [S28] a monitored failure mode, not an afterthought.

\section{Threats to Validity}\label{sec:threats}
\textbf{Construct validity.} Terminology fragments across HCI, ML, robotics, and IS (\emph{reliance}, \emph{trust calibration}, \emph{deferral}, \emph{facilitation}); relevant work under adjacent labels may be missed. We mitigated this with three concept groups, pilot calibration against known works,
terminology re-probing as new labels surfaced, and backward and forward snowballing~\cite{wohlin2014}. \textbf{Internal validity.} Screening and coding involve judgment; we used explicit criteria, a shared extraction form, and author cross-checking. 
\textbf{Interpretive validity.} Taxonomy-driven synthesis risks over-grouping heterogeneous studies; traceability from every claim to coded cells limits this. \textbf{Conclusion validity.} Positive-result bias in the underlying empirical literature (documented for medical human-AI teaming [S49]) may inflate the apparent effectiveness of individual interventions; our aggregate claims lean on the meta-analytic and reliability-based syntheses [S48, S49], which are less exposed, and the distribution counts of Table~\ref{tab:glance} describe the reviewed works rather than effect sizes. \textbf{External validity.} 
The review covers English-language, mostly peer-reviewed work (preprints and benchmarks only via flagged snowballing); the taxonomies should be read as a synthesis of published evidence. Section 8 is an illustrative instantiation as it shows that the loop can be populated end-to-end in one setting, and no claim of generality follows from it.

\section{Safe and Responsible Innovation Statement}

This review synthesizes mechanisms that sense people (affect, engagement, group dynamics) and act on them. Such loops risk surveillance, manipulation, cultural unfairness, over-reliance, and deskilling; several reviewed works quantify these harms [S8, S14, S20, S28, S48]. We therefore treat privacy-lighter modality portfolios, calibrated uncertainty communication, human-in-command facilitation, and auditable, revocable adaptation policies as first-class MCAL design elements, and flag assurance of adaptive interventions as an open requirement. The review processes no personal data.

\section{Conclusion}
We reviewed 50 works on adaptive human-AI collaboration and synthesized them into three taxonomies (multimodal context modeling up to collective states, uncertainty-aware intervention, and longitudinal co-adaptation) integrated in the MCAL dual-timescale reference
model and instantiated on a mixed human-robot workspace. The evidence locates the synergy gap not in any single component but in the unclosed loop between them; the taxonomies, model, and data are offered as a foundation for building and evaluating systems that learn how to collaborate.




\bibliographystyle{ACM-Reference-Format}
\bibliography{references}

\section*{Reviewed Works}
{\footnotesize\raggedright\sloppy\begin{description}[leftmargin=2.6em,itemsep=0pt,parsep=0pt,style=unboxed]
\item[{[S1]}] Chen, X., Yang, J.: X-Fi: A modality-invariant foundation model for multimodal human sensing. In: International Conference on Learning Representations (ICLR) (2025)
\item[{[S2]}] Liang, P.P., Goindani, A., Chafekar, T., Mathur, L., Yu, H., Salakhutdinov, R., Morency, L.-P.: HEMM: Holistic evaluation of multimodal foundation models. In: NeurIPS Datasets and Benchmarks Track (2024)
\item[{[S3]}] Mathur, L., Qian, M., Liang, P.P., Morency, L.-P.: Social Genome: Grounded social reasoning abilities of multimodal models. In: Proc. EMNLP, pp. 24868--24891 (2025)
\item[{[S4]}] Wilf, A., Mathur, L., Mathew, S., Ko, C., Kebe, G. Y., Liang, P. P., and Morency, L.-P. 2023. Social-IQ 2.0 Challenge: Benchmarking Multimodal Social Understanding. Artificial Social Intelligence Workshop/Challenge at ICCV 2023; challenge website and repository.
\item[{[S5]}] Binz, M., Akata, E., Bethge, M., et al.: A foundation model to predict and capture human cognition. Nature 644(8078), 1002--1009 (2025)
\item[{[S6]}] Zhou, P., Madaan, A., Potharaju, S.P., et al.: How far are large language models from agents with theory-of-mind? arXiv:2310.03051 (2023)
\item[{[S7]}] M\"uller, P., Balazia, M., Baur, T., et al.: MultiMediate'24: Multi-domain engagement estimation. In: Proc. 32nd ACM Int. Conf. on Multimedia, pp. 11377--11382 (2024)
\item[{[S8]}] Withanage Don, D.S., Funk, M., Balazia, M., et al.: MultiMediate'25: Cross-cultural multi-domain engagement estimation. In: Proc. 33rd ACM Int. Conf. on Multimedia, pp. 14150--14155 (2025)
\item[{[S9]}] Alipour, M., Moghaddam, M.T., Vaidhyanathan, K., Kj\ae rgaard, M.B.: Toward changing users behavior with emotion-based adaptive systems. In: Proc. 31st ACM Conf. on User Modeling, Adaptation and Personalization (UMAP) (2023)
\item[{[S10]}] Li, S., Zheng, P., Liu, S., Wang, Z., Wang, X.V., Zheng, L., Wang, L.: Proactive human-robot collaboration: Mutual-cognitive, predictable, and self-organising perspectives. Robotics and Computer-Integrated Manufacturing 81, 102510 (2023)
\item[{[S11]}] Steyvers, M., Tejeda, H., Kerrigan, G., Smyth, P.: Bayesian modeling of human--AI complementarity. PNAS 119(11), e2111547119 (2022)
\item[{[S12]}] Cresswell, J.C., Sui, Y., Kumar, B., Vouitsis, N.: Conformal prediction sets improve human decision making. In: Proc. 41st ICML, PMLR 235, pp. 9439--9457 (2024)
\item[{[S13]}] Ren, A.Z., Dixit, A., Bodrova, A., et al.: Robots that ask for help: Uncertainty alignment for large language model planners. In: Proc. 7th Conf. on Robot Learning (CoRL), PMLR 229, pp. 661--682 (2023)
\item[{[S14]}] Zhou, K., Hwang, J.D., Ren, X., Sap, M.: Relying on the unreliable: The impact of language models' reluctance to express uncertainty. In: Proc. 62nd ACL, pp. 3623--3643 (2024)
\item[{[S15]}] Zhou, K., Jurafsky, D., Hashimoto, T.: Navigating the grey area: How expressions of uncertainty and overconfidence affect language models. In: Proc. EMNLP, pp. 5506--5524 (2023)
\item[{[S16]}] Okamura, K., Yamada, S.: Adaptive trust calibration for human-AI collaboration. PLoS ONE 15(2), e0229132 (2020)
\item[{[S17]}] Jitkrittum, W., Gupta, N., Menon, A.K., Narasimhan, H., Rawat, A.S., Kumar, S.: When does confidence-based cascade deferral suffice? Advances in Neural Information Processing Systems 36: 9891-9906 (2023)
\item[{[S18]}] Westphal, M., Hemmer, P., V\"ossing, M., Schemmer, M., Vetter, S., Satzger, G.: Towards understanding AI delegation: The role of self-efficacy and visual processing ability. ACM Trans. Interactive Intelligent Systems 15(1) (2024)
\item[{[S19]}] Bansal, G., Wu, T., Zhou, J., Fok, R., Nushi, B., Kamar, E., Ribeiro, M.T., Weld, D.: Does the whole exceed its parts? The effect of AI explanations on complementary team performance. In: Proc. CHI, Article 81 (2021)
\item[{[S20]}] Bu\c{c}inca, Z., Malaya, M.B., Gajos, K.Z.: To trust or to think: Cognitive forcing functions can reduce overreliance on AI in AI-assisted decision-making. Proc. ACM Hum.-Comput. Interact. 5(CSCW1), Article 188 (2021)
\item[{[S21]}] Vasconcelos, H., J\"orke, M., Grunde-McLaughlin, M., Gerstenberg, T., Bernstein, M.S., Krishna, R.: Explanations can reduce overreliance on AI systems during decision-making. Proc. ACM Hum.-Comput. Interact. 7(CSCW1) (2023)
\item[{[S22]}] Cabrera, \'A.A., Perer, A., Hong, J.I.: Improving human-AI collaboration with descriptions of AI behavior. Proc. ACM Hum.-Comput. Interact. 7(CSCW1) (2023)
\item[{[S23]}] Jussupow, E., Spohrer, K., Heinzl, A., Gawlitza, J.: Augmenting medical diagnosis decisions? An investigation into physicians' decision-making process with artificial intelligence. Information Systems Research 32(3), 713--735 (2021)
\item[{[S24]}] Madras, D., Pitassi, T., Zemel, R.: Predict responsibly: Improving fairness and accuracy by learning to defer. In: NeurIPS (2018)
\item[{[S25]}] Mozannar, H., Sontag, D.: Consistent estimators for learning to defer to an expert. In: Proc. 37th ICML, PMLR 119, pp. 7076--7087 (2020)
\item[{[S26]}] Mozannar, H., Lang, H., Wei, D., Sattigeri, P., Das, S., Sontag, D.: Who should predict? Exact algorithms for learning to defer to humans. In: Proc. AISTATS, PMLR 206 (2023)
\item[{[S27]}] F\"ugener, A., Grahl, J., Gupta, A., Ketter, W.: Cognitive challenges in human--artificial intelligence collaboration: Investigating the path toward productive delegation. Information Systems Research 33(2), 678--696 (2022)
\item[{[S28]}] F\"ugener, A., Grahl, J., Gupta, A., Ketter, W.: Will humans-in-the-loop become borgs? Merits and pitfalls of working with AI. MIS Quarterly 45(3), 1527--1556 (2021)
\item[{[S29]}] Bu\c{c}inca, Z., Swaroop, S., Paluch, A.E., Murphy, S.A., Gajos, K.Z.: Towards optimizing human-centric objectives in AI-assisted decision-making with offline reinforcement learning. arXiv:2403.05911 (2024)
\item[{[S30]}] Todi, K., Bailly, G., Leiva, L.A., Oulasvirta, A.: Adapting user interfaces with model-based reinforcement learning. In: Proc. CHI (2021)
\item[{[S31]}] Feng, X., et al.: Large language model-based human-agent collaboration for complex task solving. In: Findings of EMNLP, pp. 1336--1357 (2024)
\item[{[S32]}] Gao, G., Taymanov, A., Salinas, E., Mineiro, P., Misra, D.: Aligning LLM agents by learning latent preference from user edits. In: NeurIPS (2024)
\item[{[S33]}] Shao, Y., Samuel, V., Jiang, Y., Yang, J., Yang, D.: Collaborative Gym: A framework for enabling and evaluating human-agent collaboration. arXiv:2412.15701 (2024)
\item[{[S34]}] Tschandl, P., Rinner, C., Apalla, Z., et al.: Human--computer collaboration for skin cancer recognition. Nature Medicine 26(8), 1229--1234 (2020)
\item[{[S35]}] Moghaddam, M.T., Santilli, T., Alipour, M.: LLM-assisted reinforcement learning for affective game adaptation. Proc. ACM Hum.-Comput. Interact. 10(4), Article EICS029, 1-30 (2026). doi:10.1145/3816781
\item[{[S36]}] Kim, S., Eun, J., Oh, C., Suh, B., Lee, J.: Bot in the bunch: Facilitating group chat discussion by improving efficiency and participation with a chatbot. In: Proc. CHI (2020)
\item[{[S37]}] Kim, S., Eun, J., Seering, J., Lee, J.: Moderator chatbot for deliberative discussion: Effects of discussion structure and discussant facilitation. Proc. ACM Hum.-Comput. Interact. 5(CSCW1) (2021)
\item[{[S38]}] Lee, S.C., Song, J., Ko, E.-Y., Park, S., Kim, J., Kim, J.: SolutionChat: Real-time moderator support for chat-based structured discussion. In: Proc. CHI (2020)
\item[{[S39]}] Tessler, M.H., Bakker, M.A., Jarrett, D., et al.: AI can help humans find common ground in democratic deliberation. Science 386(6719), eadq2852 (2024)
\item[{[S40]}] Traeger, M.L., Strohkorb Sebo, S., Jung, M., Scassellati, B., Christakis, N.A.: Vulnerable robots positively shape human conversational dynamics in a human-robot team. PNAS 117(12), 6370--6375 (2020)
\item[{[S41]}] Tennent, H., Shen, S., Jung, M.: Micbot: A peripheral robotic object to shape conversational dynamics and team performance. In: Proc. 14th ACM/IEEE HRI, pp. 133--142 (2019)
\item[{[S42]}] van Zoelen, E.M., van den Bosch, K., Neerincx, M.: Becoming team members: Identifying interaction patterns of mutual adaptation for human-robot co-learning. Frontiers in Robotics and AI 8, 692811 (2021)
\item[{[S43]}] Te'eni, D., Yahav, I., Zagalsky, A., Schwartz, D.G., Silverman, G., Cohen, D., Mann, Y.,
Lewinsky, D.: Reciprocal human-machine learning: A theory and an instantiation for the case of
message classification. Management Science 72(1), 167–192 (2026). doi:10.1287/mnsc.2022.03518
\item[{[S44]}] Nwankwo, L., Ellensohn, B., Rauch, C., Rueckert, E.: SIL: Symbiotic interactive learning for language-conditioned human-agent co-adaptation. In: Proc. 35th IEEE RO-MAN (2026)
\item[{[S45]}] Wang, W., Gao, G., Agarwal, R.: Friend or foe? Teaming between artificial intelligence and workers with variation in experience. Management Science 70(9), 5753--5775 (2024)
\item[{[S46]}] Pedreschi, D., Pappalardo, L., Ferragina, E., et al.: Human-AI coevolution. Artificial Intelligence 339, 104244 (2025)
\item[{[S47]}] Hemmer, P., Schemmer, M., K\"uhl, N., V\"ossing, M., Satzger, G.: Complementarity in human-AI collaboration: Concept, sources, and evidence. European Journal of Information Systems 34, 6, 979–1002 (2025). doi:10.1080/0960085X.2025.2475962
\item[{[S48]}] Vaccaro, M., Almaatouq, A., Malone, T.: When combinations of humans and AI are useful: A systematic review and meta-analysis. Nature Human Behaviour 8(12), 2293--2303 (2024)
\item[{[S49]}] Liu, P., Zhang, J., Chen, S., et al.: Human-AI teaming in healthcare: 1+1>2? npj Artificial Intelligence 1, 47 (2025)
\item[{[S50]}] Yin, J., Ngiam, K.Y., Tan, S.S.-L., Teo, H.H.: Designing AI-based work processes: How the
timing of AI advice affects diagnostic decision making. Management Science 71(11), 9361–9383
(2025). 
\end{description}}

\end{document}